\documentstyle[aps, manuscript]{revtex}
\tightenlines
\draft
\begin{document}
\title{Origin of FRW cosmology in slow-roll inflation
from noncompact Kaluza-Klein theory}
\author{$^1$Edgar Madriz Aguilar\footnote{
E-mail address: edgar@itzel.ifm.umich.mx}
and $^{2}$Mauricio Bellini\footnote{
E-mail address: mbellini@mdp.edu.ar}}
\address{$^1$Instituto de F\'{\i}sica y Matem\'aticas,
AP: 2-82, (58040) Universidad Michoacana de San Nicol\'as de Hidalgo,
Morelia, Michoac\'an, M\'exico.\\
$^2$ Consejo Nacional de Investigaciones Cient\'{\i}ficas y
T\'ecnicas (CONICET) \\
and\\
Departamento de F\'{\i}sica, Facultad de Ciencias Exactas y Naturales,
Universidad Nacional de Mar del Plata,
Funes 3350, (7600) Mar del Plata, Argentina.}

\vskip .5cm
\maketitle
\begin{abstract}
Using a recently introduced formalism
we discuss slow-roll inflation
from Kaluza-Klein theory without
the cylinder condition. In particular, some examples corresponding
to polynomic and hyperbolic $\phi$-potentials
are studied.
We find that the evolution of the fifth coordinate
should be determinant for both,
the evolution of the early inflationary
universe and the quantum fluctuations.
\end{abstract}
\vskip .2cm                             
\noindent
Pacs numbers: 04.20.Jb, 11.10.kk, 98.80.Cq \\
\vskip 1cm
\section{Introduction}

The possibility that our universe es embedded in a higher dimensional
space has generated a great deal of active interest.
In Brane-World (BW)\cite{au,aau,bu}
and Space-Time-Matter (STM)\cite{WE}
theories the usual constraint on Kaluza-Klein
(KK) models, namely the cylinder condition, is relaxed so the extra dimensions
are not restricted to be compact.
An alternative idea to geometrical
compactification has been explored by Randall and Sundrum\cite{rs},
who demonstrated that for sufficiently low energies the probability
of losing energy to the KK states is very small.
The first important question concerning solutions in 5D is to check
whether they give back the standard 4D results. In particular, cosmological
models should be developed from a Friedmann-Robertson-Walker (FRW) cosmology.

Inflation has nowadays become a standard ingredient for the
description of the early Universe.
In fact, it solves some of the problems of the standard big-bang scenario and
also makes predictions about cosmic microwave background radiation
(CMBR) anisotropies which are being measured with
higher and higher precision.
The first model of inflation was proposed by A. Starobinsky in 1979\cite{1'}.
A much simpler inflationary model with a clear motivation was developed by
Guth in the 80's\cite{Guth},
in order to solve some of the shortcomings of the big bang theory, and in
particular, to explain the extraordinary homogeneity of the observable 
universe. However, the universe after inflation in this scenario
becomes very inhomogeneous. Following a detailed investigation
of this problem, A. Guth and E. Weinberg concluded that the old
inflationry model could not be improved\cite{GW}. 
These problems were sorted out by A. Linde in 1983
with the introduction of chaotic inflation\cite{lin}.
In this scenario inflation can occur in theories with
potentials such as $V(\phi) \sim \phi^n$. It may
begin in the absence of thermal equilibrium in the early
universe, and it may start even at the Planckian density, in
which case the problem of initial conditions for inflation
can be easily solved\cite{libro}.

Recently there has been significant progress made in the field of string
cosmology\cite{x,xx}. Arguably, among the greatest triumphs are the string
realizations of the inflationary universe paradigm\cite{xxx}.
Moreover, inflationary cosmlogy from STM models
has been subject of great interest in the last years\cite{wesson,cqg,NPB}.
In a novel approach recently developed
\cite{PLB}
was suggested that 
the evolution of the
early universe could be described by a geodesic trayectory of a
5D metric, so that the effective 4D
FRW background metric should be a hypersurface on
a constant fifth dimension.
In this paper we extend this approach to other inflationary
models. The work is organized as follows: in sect. II the formalism
proposed in\cite{PLB} is reviewed and extended. In sect. III we
study inflationary dynamics taking into account
the slow-roll conditions.
Along that section inflationary models becoming
from polynomic (quadratic and quartic) and hyperbolic $\phi$-
potentials are discussed. Finally, in sect. IV we develop some
final comments.

\section{Formalism reviewed and extended}

We consider the 5D metric, recently introduced by Ledesma and
Bellini (LB)\cite{PLB}
\begin{equation}\label{6}
dS^2 = \psi^2 dN^2 - \psi^2 e^{2N} dr^2 - d\psi^2,
\end{equation}
where the parameters ($N$,$r$)
are dimensionless and the fifth coordinate
$\psi $ has spatial unities.
The metric (\ref{6}) describes a
flat 5D manifold in apparent vacuum ($G_{AB}=0$).
Furthermore, 
on hypersurfaces $\psi=$ const. the 4D induced pressure (${\rm p}$)
and energy density ($\rho$) are given by
\begin{equation}
{\rm p} = - \frac{3}{8\pi G\psi^2}, \qquad \rho=\frac{3}{8\pi G\psi^2},
\end{equation}
such that the 4D equation of state on hypersurfaces with constant
$\psi$ is ${\rm p}=-\rho$, which corresponds with a vacuum. Hence,
systems in an apparent vacuum
on hypersurfaces with constant $\psi$ in a 5D manifold
described by eq. (\ref{6}), complies with a 4D vacuum equation of state.
In particular, the case for which the squared Hubble
parameter is given by $H^2_0 = \Lambda/(3\psi^2)$ represents a de
Sitter expansion\cite{pdl} governed by the cosmological constant $\Lambda$.

As in the paper \cite{PLB} we shall consider
the case 
where $N$ only depends on the cosmic time $t$: $N=N(t)$.
The relevant Christoffel symbols for the geodesic of the metric
(\ref{6}) in a comoving frame  $U^r = {\partial r/\partial S}=0$, are
\begin{equation}
\Gamma^N_{\psi\psi}= 0, \quad \Gamma^N_{\psi N}= 1/\psi, \quad
\Gamma^{\psi}_{NN}= \psi, \quad \Gamma^{\psi}_{N \psi}= 0,
\end{equation}
so that the geodesic dynamics ${dU^C \over dS} = \Gamma^C_{AB} U^A U^B$
is described by the following equations of motion
for the velocities $U^A$
\begin{eqnarray}
&& \frac{dU^{N}}{dS} = -\frac{2}{\psi} U^N U^{\psi}, \\
&& \frac{dU^{\psi}}{dS} = -\psi U^N U^N, \\
&&  \psi^2 U^N U^N - U^{\psi} U^{\psi} =1, \label{geo}
\end{eqnarray}
where the eq. (\ref{geo}) describes the constraint condition
$g_{AB}U^A U^B=1$.
From the general solution $\psi U^N  = {\rm cosh}[S(N)]$,
$U^{\psi}=-{\rm sinh}[S(N)]$, we obtain the equation
that describes the geodesic evolution for $\psi$
\begin{equation}
\frac{d\psi}{dN} = \frac{U^{\psi}}{U^N} = -\psi {\rm tanh}[S(N)],
\end{equation}
where $S(N)=-N$ gives the number of e-folds of the universe.
If we take ${\rm tanh}[S(N)]
=-1/u(N)$,
we obtain
\begin{equation}\label{21}
\psi(N) = \psi_0 e^{\int dN/u(N)},
\end{equation}
for the velocities
\begin{equation}
U^{\psi} = - \frac{1}{\sqrt{u^2(N)-1}}, \qquad
U^N=\frac{u(N)}{\psi\sqrt{u^2(N)-1}},
\end{equation}
where $\psi_0$ in eq. (\ref{21}) is a constant of integration.
As in a previous paper\cite{PLB}, we are interested in the case
in which $\psi =H^{-1}$, where $H$ is the classical
Hubble parameter. With this choice the constant $\psi_0 = H^{-1}_0$
describes the primordial Hubble horizon, which should be of the
order of the Planckian length.
Furthermore, the function $u$ is given by $u(N)=-{H \over dH/dN} >0$,
because $dH/dN <0$ during inflation.

The resulting 5D metric is given by
\begin{equation}\label{4d}
dS^2=dt^2-e^{2\int H(t)dt} dR^2-dL^2,
\end{equation}
with $t=\int \psi(N) dN$, $R=r\psi$ and $L=\psi_0$.
With this representation, we obtain the following new
velocities $ \hat U^A ={\partial \hat x^A \over \partial x^B} U^B$
\begin{equation} \label{10}
U^t=\frac{2u(t)}{\sqrt{u^2(t)-1}}, \qquad
U^R=\frac{-2r}{\sqrt{u^2(t)-1}}, \qquad U^L=0,
\end{equation}
where
the old velocities $U^B$ are $U^N$, $U^r=0$ and $U^{\psi}$.
Furthermore, the velocities $\hat U^B$ complies with the
constraint condition
\begin{equation}\label{con}
\hat g_{AB} \hat U^A \hat U^B =1.
\end{equation}
The important fact here is that the new frame give us an effective
spatially flat FRW metric embedded in a 5D manifold where the
fifth coordinate $L=\psi_0$ is the primordial Hubble horizon,
which emerges naturally as a constant in this representation.

The solution $N={\rm arctanh}[1/u(t)]$ corresponds to a power-law
expanding universe with time dependent power $p(t)$
for a scale factor $a \sim t^{p(t)}$. Since $H(t) = \dot a/a$, the
resulting Hubble parameter is
\begin{equation}\label{h}
H(t)=\dot p {\rm ln}(t/t_0) +p(t)/t,
\end{equation}
where $t_0$ is the initial time.
The function $u$ written as a function of time is
\begin{equation}
u(t) = -\frac{H^2}{\dot H},
\end{equation}
where the dot represents the derivative with respect to the time.

With this representation the universe can be viewed as born
in a state with $S\simeq 0$ (i.e., in a 4D vacuum state ${\rm p}\simeq-\rho$),
where the fifth coordinate is given by the Hubble horizon
in a comoving
frame $dr=0$, such that the effective 4D spacetime is a FRW metric
\begin{equation}\label{4d1}
dS^2 = dt^2 - e^{2\int H(t) dt} dR^2 - dL^2 \rightarrow
ds^2 = dt^2 - e^{2\int H(t) dt} dR^2.
\end{equation}
In this framework we can define
the $5D$ lagrangian
\begin{equation}\label{l1}
{\cal L}(\phi,\phi_{,A}) = -\sqrt{-^{(5)}g} \left[
\frac{1}{2} g^{AB} \phi_{,A} \phi_{,B} + V(\phi)\right],
\end{equation}
for the scalar field $\phi(N,r,\psi)$ with the metric (\ref{6}).
Here, $^{(5)}g$ is the determinant of the 5D metric tensor in
(\ref{6}) and $V(\phi)$
is the potential. On the geodesic $N={\rm arctanh}[1/u(t)]$ in the comoving
frame $dr=0$, the effective 4D lagrangian for the metric (\ref{4d1}) is
\begin{equation}
{\cal L}\left(\phi,\phi_{,A}\right)
\rightarrow {\cal L}\left(\phi,\phi_{,\mu}\right) =
-\sqrt{-^{(4)} g} \left[ \frac{1}{2} g^{\mu\nu} \phi_{,\mu}
\phi_{,\nu} +V(\phi)\right],
\end{equation}
where $^{(4)}g $ is the determinant of the metric tensor in the
4D effective FRW background metric (\ref{4d1})
and $\phi(t,R,L)\equiv \phi(t,R)$.
In this frame, the 4D energy 
density and the pressure are\cite{PLB}
\begin{eqnarray}
&& 8 \pi G \rho = 3 H^2,\\
&& 8\pi G {\rm p} = -(3H^2 + 2 \dot H),
\end{eqnarray}
with $H(t) =\dot a/a$ for a given
scale factor $a(t) \sim t^{p(t)}$ and $L=\psi_0$ is
of the order of the Planckian length.
As was emphatized in a previous work\cite{PLB} such that formalism
can be sucessfully applied to many inflationary models.
With the aim to ilustrate it, in the following section we shall
study some inflationary models.
$V(\phi) \sim \phi^n$.

\section{Slow-roll inflation}

The dynamics of the inflaton field during inflation is characterized by
the equations
\begin{eqnarray}
&& \ddot\phi + 3 H \dot\phi + V'(\phi) =0, \label{a}\\
&& \dot\phi = - \frac{M^2_p}{4\pi} H'(\phi),\label{b}
\end{eqnarray}
where the prime denotes the derivative with
respect to $\phi$. If the slow-roll conditions\cite{liddle}
are fulfilled,
\begin{equation}\label{sl}
\gamma = \frac{M^2_p}{4\pi} \left(\frac{H'}{H}\right)^2 \ll1, \qquad
\eta = \frac{M^2_p}{4\pi} \frac{H''}{H} \ll 1,
\end{equation}
the Friedmann equation can be approximated to
\begin{equation}
H^2 \simeq \frac{8\pi}{3M^2_p} V(\phi),
\end{equation}
such that $H' \simeq  {\sqrt{4\pi} \over 3 \sqrt{2} M_p} {V' \over V^{1/2}}$,
and the eq. (\ref{b}) can be approximated to
\begin{equation}\label{c}
\dot\phi \simeq - \frac{V'}{3 H} = -\frac{M_p}{\sqrt{4\pi}} \frac{V'}{V^{1/2}}.
\end{equation}
In other words, slow-roll conditions implicate that
\begin{equation}          \label{alpha}
\frac{3}{8\pi G \psi^2} \simeq V(\phi),
\end{equation}
where $G=M^{-2}_p$ is the gravitational constant
and $M_p = 1.2 \  10^{19}$ GeV is the Planckian mass.
This is a very important expression that say us that
the evolution of the universe [governed by $V(\phi)$], has its origin
in the evolution of the fifth dimension $\psi=H^{-1}$.
The expression (\ref{alpha}) is approximately fulfilled
in the early universe before the inflaton field begins to oscillate
around the minimum of the potential $V(\phi)$.
Note that the function $u$, written as a function of $\phi$
\begin{equation}\label{12}
u(\phi) = \frac{4\pi}{M^2_p} \left(\frac{H}{H'}\right)^2,
\end{equation}
which
is exactly the inverse of the slow-roll parameter $\gamma$ [see the first
eq. in (\ref{sl})]: $u = 1/\gamma$. Hence, the condition $u \gg 1$ is
guarranted during inflation. This is a very general result valid to all
the models of inflation. Furthermore the condition $u \gg 1$ say us
that the velocities (\ref{10}) are real. They can
be written
in terms of the Hubble parameter
\begin{equation}\label{vel}
U^t = \frac{2\left(\frac{H}{H'}\right)^2}{\sqrt{\left(
\frac{H}{H'}\right)^4 - \frac{M^4_p}{16\pi^2}}}, \qquad
U^R = -\frac{2r}{\sqrt{
16\pi^2 \left(\frac{H}{H'M_p}\right)^4 -1}}, \qquad U^L=0.
\end{equation}
Furthermore, since $u \gg 1$ during inflation, we can approximate
approximate the velocities (\ref{vel})
\begin{equation}
U^t \simeq 2, \qquad U^R \simeq -2 r \gamma, \qquad U^L=0.
\end{equation}
The constraint condition (\ref{con}) implies that
(for $\gamma_0 ={\sqrt{3}\over 2r}$)
\begin{equation}\label{con1}
\gamma \simeq \gamma_0 \frac{a_0}{a(t)} \ll 1,
\end{equation}
which holds to all $t\ge t_0$ and $r \gg 1$.
Note that eq. (\ref{con1}) has its origin in a geometrical property;
the condition (\ref{con}).
The constraint condition (\ref{con1}) can be written as
\begin{equation}
u(t) a_0 \simeq u_0 a(t),
\end{equation}
where $u_0=r/\sqrt{3} \gg 1$ and $u(t) = -H^2/\dot H$.

Potentials like
\begin{equation}\label{aa}
V(\phi) = \frac{\lambda}{ n} \phi^n,
\end{equation}
are interesting for inflationary cosmology.
For these potentials the eq. (\ref{c}) becomes
\begin{equation}
\dot\phi = - \frac{M_p}{\sqrt{4\pi}} \sqrt{n\lambda} \phi^{\frac{n-2}{2}},
\end{equation}
with solutions
\begin{eqnarray}
&& \phi^{\frac{4-n}{2}}(t) =
\phi^{\frac{4-n}{2}}_0 - \frac{M_p \sqrt{n\lambda}
(4-n)}{4\sqrt{\pi}}t. \label{n} \\
&& \phi(t) = \phi_0 e^{-\sqrt{\frac{\lambda}{6\pi}} t},\label{nn}
\end{eqnarray}
for $n\neq 4$ and $n=4$, respectively.
Here, $\phi_0$ is the initial value of the scalar field: $\phi_0\equiv
\phi(t_0) \geq \phi(t)$ for symmetric potentials $V(\phi) = V(-\phi)$,
such that $V(\phi_0) \simeq M^4_p$.
For any $n$, the scale factor $a(t) = a_0 \left(t/t_0\right)^{p(t)}$,
can be written as a function of $\phi(t)$
\begin{equation} \label{aaa}
\frac{a(t)}{a_0} = e^{\frac{2\pi}{nM^2_p} \sqrt{2/3} \left[\phi^2_0 -
\phi^2(t)\right]}.
\end{equation}
During inflaton the amplitude of the quantum fluctuations are of the order
of the Hubble parameter: $|\delta\phi| \simeq {H\over 2\pi}$\cite{alpha,libro},
so that
\begin{equation}
|\delta\phi| \simeq \frac{1}{2\pi\psi},
\end{equation}
$|\delta\phi|$ is dominatted by the evolution of the fifth coordinate.
Furthermore, inflation ends when the inflaton field assume the value
$\phi_e \simeq {n M_p \over 4\sqrt{3\pi}}$.
Since $H=\dot a/a$, from the eqs. (\ref{h}) and (\ref{aaa}) 
we obtain the temporal dependence
of the time-dependent power $p(t)$ [written as a function of $\phi(t)$]
\begin{equation}
p(t) = \frac{\frac{2\pi}{nM^2_p} \sqrt{2/3} \left[\phi^2_0 -
\phi^2(t)\right]}{{\rm ln}\left[\frac{t}{t_0}\right]},
\end{equation}
and, with more generality
\begin{equation}\label{27}
p(t) = \frac{N(t)}{{\rm ln}\left[\frac{t}{t_0}\right]},
\end{equation}
where $N(t)={\Large\int}^{t}_{t_0} H(t') dt'$ is the number of e-folds
from $t_0$ to $t$.

In order to ilustrate the generality of the formalism here studied,
in the next section we
shall develop some particular inflationary examples described
by symmetric $\phi$-potentials.

\section{Some Examples}

Slow-roll inflation is well described by symmetric potentials
like (\ref{aa}) or $V(\phi) \sim {\rm sinh}^2(\beta\phi)$.
In this section we shall study the dynamics of slow-roll inflation
in quadratic, quartic and hyperbolic $\phi$-potentials.

\subsection{Massive scalar field}

As a first example we consider a massive scalar field described by a
quadratic potential $V(\phi) = {m^2\over 2} \phi^2$. This case corresponds
to $\lambda=m^2$ and $n=2$ in the equation (\ref{aa}).
Its temporal dependence is described by the eq. (\ref{n}), so that
\begin{equation}
\phi(t) = \phi_0 - \frac{M_p m}{\sqrt{2 \pi}} t,
\end{equation}
where $m$ is the mass of the inflaton field. The Hubble parameter
can be written as a function of $\phi$
\begin{equation}
H(\phi) \simeq 2 \sqrt{\frac{\pi}{3}} \frac{m}{M_p} \phi,
\end{equation}
and $p(t)$ can be obtained from eq. (\ref{27})
\begin{equation}
p(t) = \sqrt{\frac{2}{3}} \pi \frac{m}{M_p} \frac{t}{{\rm ln}(t/t_0)}
\left( \sqrt{\frac{2}{\pi}} \phi_0 - \frac{m M_p}{2\pi} t\right),
\end{equation}
where $\phi_0 > {1\over 2} \sqrt{{1\over 2\pi}} m M_p t_e$ ($t_e$
is the time at the end of inflation), due to the
fact $p>0$ in an expanding universe. In particular, $p>1$ during the
inflationary epoch. Note that $p$ decreases with time during inflation.
Thus, at the begining $p$ should take a value very large that decreases
until values very close to $p\simeq 1$ at the end of inflation.
The function $u$ [see eq. (\ref{12})], can be written as a function of the
inflaton field
\begin{equation}
u(\phi) = \frac{4\pi}{M^2_p} \phi^2,
\end{equation}
which is $u \gg 1$ because during inflation the slow-roll conditions
implicate
\begin{equation}
\phi^2 \gg \frac{M^2_p}{4\pi}.
\end{equation}
The velocities (\ref{vel}) can be written as a function of the inflaton
field
\begin{equation}
U^t = \frac{2}{\sqrt{1-\frac{M^4_p}{16\pi^2\phi^4}}}
\simeq 2,
\qquad U^R = \frac{-2r}{\sqrt{\frac{16\pi^2\phi^4}{M^4_p}-1}}
\simeq -2 r\gamma(\phi).
\end{equation}
The number of e-folds and the
evolution of the fifth coordinate can be written as a function
of the inflaton field
\begin{eqnarray}
&& N(\phi) = \frac{\pi}{2 M^2_p} \sqrt{\frac{2}{3}} \left(\phi^2_0 -
\phi^2\right), \\
&& \psi(\phi) = \frac{1}{2} \sqrt{\frac{3}{\pi}} \frac{M_p}{m}\phi^{-1},
\end{eqnarray}
where $\phi_0 = {\sqrt{2}\over m} M^2_p$ is the initial value
of the inflaton field, ${M_p \over 2\sqrt{3\pi}} < \phi <\phi_0$ and the initial value of
the fifth coordinate $\psi_0=H^{-1}_0$ is of the order of the Planckian
length
\begin{equation}\label{mas}
\psi_0 \simeq \frac{1}{2} \sqrt{\frac{3}{2\pi}} \frac{1}{M_p}.
\end{equation}

\subsection{Self interacting scalar field}

Other interesting example that describes a self interacting scalar
field is given by a quartic potential $V(\phi) = {\lambda\over 4} \phi^4$.
Here, $\lambda \ll 1$ is a dimensionless constant 
and the classical evolution for the inflaton field
is given by the eq. (\ref{nn}).
The Hubble parameter is
\begin{equation}
H(\phi) \simeq \frac{1}{M_p} \sqrt{\frac{2\pi \lambda}{3}} \phi^2,
\end{equation}
and the temporal evolution for $\phi(t)$ is given by eq. (\ref{nn}).
Hence, the temporal dependence for $p(t)$ is
\begin{equation}
p(t) =\frac{\phi^2_0}{\sqrt{6} M^2_p {\rm ln}(t/t_0)} \left(
1- e^{-2\sqrt{\frac{\lambda}{6\pi}} t}\right),
\end{equation}
for $t>M^{-1}_p$. Note that $p$ decresases with time.
The function $u$ is given by
\begin{equation}
u(\phi) \simeq \frac{\phi^2}{M^2_p} \gg 1,
\end{equation}
such that $\phi^2 \gg M^2_p$.
The velocities (\ref{vel}) for a model $V(\phi)={\lambda \over 4}
\phi^4$ become
\begin{equation}
U^t =\frac{2}{\sqrt{1-\frac{M^4_p}{\phi^4}}}
\simeq 2, \qquad U^R = \frac{-2r}{\sqrt{\frac{\phi^4}{M^4_p}-1}}
\simeq -2 r\gamma(\phi).
\end{equation}
On the other hand,
the number of e-folds and the
evolution of the fifth coordinate can be related to
the inflaton field
\begin{eqnarray}
&& N(\phi) = \frac{\pi}{4 M^2_p} \sqrt{\frac{2}{3}} \left(\phi^2_0 -
\phi^2\right), \\
&& \psi(\phi) = M_p \sqrt{\frac{3}{2\pi\lambda}}\phi^{-2},
\end{eqnarray}
where $\phi_0 = \left({4\over \lambda}\right)^{1/4} M_p > \phi
> {M_p \over \sqrt{3\pi}}$ and
$\psi_0 \simeq {1 \over 2} \sqrt{{3\over 2\pi}} {1\over M_p}$,
which is exactly the
value we found for a massive scalar field [see eq. (\ref{mas})].

\subsection{Hyperbolic potential}

As a third example we consider a symmetric potential given by
\begin{equation}
V(\phi) = V_0 \  {\rm sinh}^2(\beta\phi),
\end{equation}
where the parameter $\beta >0$ has dimensions of the inverse of the mass.
The dynamics of the inflaton field is given by
\begin{equation}
\dot\phi \simeq -\frac{2 V^{1/2}_0 M_p \beta}{\sqrt{24\pi}} {\rm cosh}(\beta\phi),
\end{equation}
so that the temporal evolution of the inflaton field yields
\begin{equation}\label{if}
\phi(t) = \phi_0 - {\rm ln}\left\{ {\rm tan}\left[
-\frac{\sqrt{24\pi}}{\beta^2 M^3_p} t\right]\right\}.
\end{equation}
The Hubble parameter is
\begin{equation}
H(\phi) \simeq \sqrt{\frac{8\pi}{3}}
\frac{V^{1/2}_0}{M_p} {\rm sinh}(\beta\phi),
\end{equation}
so that the number of e-folds is
\begin{equation}
N[\phi(t)] =
\frac{1}{2\beta} \left\{{\rm ln}\left[{\rm tanh}\left(\beta\phi(t)\right)
-1\right]+ {\rm ln}\left[{\rm tanh}\left(\beta\phi(t)\right)
+1\right]\right\}^{\phi_0}_{\phi(t)},
\end{equation}
where $\phi(t)$ is given by (\ref{if}).
The slow-roll parameter $\gamma$ [see eq. (\ref{sl})] for this model
is
\begin{equation}
\gamma(\phi) \simeq \frac{M^2_p}{4\pi} \beta^2 {\rm coth}^2(\beta\phi),
\end{equation}
which complies with the slow-roll conditions for $\beta\phi_0 \simeq 1$
for $\beta \ll M^{-1}_p$. Note that, as in the other examples,
slow-roll conditions implies
that $\phi$ must take transplanckian values.
Furthermore,
the function $u\gg 1$  for this inflationary model is given
by
\begin{equation}
u(\phi) \simeq \frac{4\pi}{M^2_p \beta^2} {\rm tanh}^2(\beta\phi).
\end{equation}
Finally, the evolution of the fifth coordinate can be written as
a function of $\phi$
\begin{equation}
\psi(\phi) \simeq \sqrt{\frac{3}{8\pi}} \frac{M_p}{V^{1/2}_0} {\rm sech}
(\beta\phi),
\end{equation}
Such that the value of the fifth corrdinate in the metric (\ref{10})
should be $\psi_0\simeq {1\over 2} \sqrt{{3\over 2\pi}} {1\over M_p}
{\rm sech}(\beta\phi_0) 
\lesssim {1\over 2}\sqrt{{3\over 2\pi}} {1\over M_p}$, which
agree quite well with the value obtained in the other examples.
To finalize, in this case the evolution of the
power-law $p(t)$ for the scale factor, will be given by eq. (\ref{27})
\begin{equation}
p(t) =
\frac{1}{2\beta {\rm ln}(t/t_0)} \left\{{\rm ln}\left[{\rm tanh}\left(\beta\phi(t)\right)
-1\right]+ {\rm ln}\left[{\rm tanh}\left(\beta\phi(t)\right)
+1\right]\right\}^{\phi_0}_{\phi(t)},
\end{equation}
which decreases during the inflationary stage.

\section{Final comments}

We have studied inflationary dynamics from noncompact
KK theory.
With this representation the universe can be viewed as born
in a state with $S\simeq 0$ (i.e., in a 4D vacuum state ${\rm p}\simeq-\rho$),
where the fifth coordinate is given by the Hubble horizon
in a comoving
frame $dr=0$, such that the effective 4D spacetime is a FRW metric
\begin{displaymath}
dS^2 = dt^2 - e^{2\int H(t) dt} dR^2 - dL^2 \rightarrow
ds^2 = dt^2 - e^{2\int H(t) dt} dR^2.
\end{displaymath}
In this frame, the effective 4D energy 
density and the pressure are
\begin{displaymath}
8 \pi G \rho = 3 H^2, \qquad
8\pi G {\rm p} = -(3H^2 + 2 \dot H).
\end{displaymath}
In particular, slow-roll inflation was discussed for
$\phi$-symmetric polynomial (quadratic and quartic)
and hyperbolic potentials.
Note that the initial length $\psi_0$ we found corresponds to
the primordial Hubble horizon $H^{-1}_0$. In all the cases here
studied its value becomes below (but of the order of)
the Planckian length:
$\psi_0 \lesssim {1 \over 2} \sqrt{{3\over 2\pi}} {1\over M_p} \simeq
0.346 \times 10^{-34}$ meters.
Thus, $L \lesssim 0.346 \times 10^{-34}$ meters should be the
value of the spatial fifth coordinate in the metric (\ref{4d1}).
Other remarkeable result of this paper
resides in that the dynamics in
slow-roll inflation
is governed by the evolution of the fifth coordinate $\psi=H^{-1}$
through the geodesic $N={\rm arctan}(1/u)$, such
that expression: ${3\over 8\pi G\psi^2} \simeq V(\phi)$ is fulfilled.
Futhermore, the quantum fluctuations are also given by $\psi$:
$|\delta\phi| \simeq {1\over 2\pi \psi}$.
Finally, the formalism could be extented to models with constant or
variable cosmological parameters $\Lambda$.
However, this topic go beyond the scope of this paper.

\vskip .2cm
\centerline{\bf{Acknowledgements}}
\vskip .2cm
EMA acknowledges CONACyT and IFM of UMSNH (M\'exico)
for financial support.
MB acknowledges CONICET, AGENCIA 
and UNMdP (Argentina)
for financial support.\\

\end{document}